\documentclass[%
 aip,
 amsmath,amssymb,
reprint,%
]{revtex4-1}

\usepackage{graphicx}
\usepackage{dcolumn}
\usepackage{bm}

\usepackage[utf8]{inputenc}
\usepackage[T1]{fontenc}
\usepackage{mathptmx}
\usepackage{xcolor}

\begin{document}

\preprint{AIP/123-QED}

\title[]{Competition between shear and biaxial extensional viscous dissipation in the expansion dynamics of Newtonian and rheo-thinning liquid sheets}

\author{Ameur Louhichi}
 \altaffiliation[]{These two authors contributed equally.}
\affiliation{Laboratoire Charles Coulomb (L2C), Universit\'e de Montpellier, CNRS, Montpellier, France}%
\affiliation{Institute of Electronic Structure and Laser, FORTH, Heraklion 70013, Crete, Greece and Department of Materials Science and Technology, University of Crete, Heraklion 70013, Crete, Greece}

\author{Carole-Ann Charles}%
 \altaffiliation[]{These two authors contributed equally.}
\affiliation{Laboratoire Charles Coulomb (L2C), Universit\'e de Montpellier, CNRS, Montpellier, France}%

 \author{Srishti Arora}
 \affiliation{Laboratoire Charles Coulomb (L2C), Universit\'e de Montpellier, CNRS, Montpellier, France}%
 \author{Laurent Bouteiller}
 \affiliation{Sorbonne Universit\'e, CNRS, IPCM, Equipe Chimie des Polym\`eres, Paris, 75005, France}
 \author{Dimitris Vlassopoulos}
 \affiliation{Institute of Electronic Structure and Laser, FORTH, Heraklion 70013, Crete, Greece and Department of Materials Science and Technology, University of Crete, Heraklion 70013, Crete, Greece}
 \author{Laurence Ramos}
  \email{laurence.ramos@umontpellier.fr}
  \affiliation{Laboratoire Charles Coulomb (L2C), Universit\'e de Montpellier, CNRS, Montpellier, France}%
  \author{Christian Ligoure}
  \email{christian.ligoure@umontpellier.fr}
\affiliation{Laboratoire Charles Coulomb (L2C), Universit\'e de Montpellier, CNRS, Montpellier, France}%

\date{\today}

\begin{abstract}
When a drop of fluid hits a small solid target of comparable size, it expands radially until reaching a maximum diameter and subsequently recedes. In this work, we show that the expansion process of liquid sheets is controlled by a combination of shear (on the target) and biaxial extensional (in the air) deformations. We propose an approach toward a rational description of the phenomenon for Newtonian and viscoelastic fluids by evaluating the viscous dissipation due to shear and extensional deformations, yielding a prediction of the maximum expansion factor of the sheet as a function of the relevant viscosity. For Newtonian systems, biaxial extensional and shear viscous dissipation are of the same order of magnitude. On the contrary, for thinning solutions of supramolecular polymers, shear dissipation is negligible compared to biaxial extensional dissipation and the biaxial thinning extensional viscosity is the appropriate quantity to describe the maximum expansion of the sheets. Moreover, we show that the rate-dependent biaxial extensional viscosities deduced from drop impact experiments are in good quantitative agreement with previous experimental data and theoretical predictions for various viscoelastic liquids. 
\end{abstract}

\maketitle

\section{\label{sec:level1}Introduction}

Drop impact process on solid surfaces is an extremely active research area because of the development of high-speed imaging technology \citep{Josserand2016}, and its implication in many industrial applications, such as the impact of pesticides drops on plant leaves~\citep{Wirth1991}, or in ink jet printing \citep{deGans2004}.

We can distinguish two main experimental configurations. On the one hand, drops
are impacted on a flat solid surface with a size much larger than the drop size, in a way that the whole expansion event occurs on the target~\citep{Crooks2000,Roux2004,Ukiwe2005,Thoroddsen2008,German2009,Izbassarov2016,
Cooper-White2002}. The sheet spreads thus in intimate contact with a uniform surface. Hence, viscous dissipation is mainly shear-induced. In that case, it has been reported that the shear rate-dependent viscosity is the pertinent parameter to describe the dissipation process in the expansion dynamics of non-Newtonian fluids~\citep{Cooper-White2002,German2009,An2012,Laan2014,Andrade2015,Boyer2016}. Note, however, that the retraction dynamics was suggested to be dominated by the non linear extensional viscosity for a drop of fluid containing polymer additives \citep{Bergeron2000a}. 

The other distinguishable configuration, albeit sparsely studied, involved impacting drops on repellent surfaces where there is no contact with the surface, hence, eliminating shear dissipation. Such surfaces may include superhydrophobic surfaces~\citep{Richard2002,Bird:2013ke,Khojasteh:2016bw,Martouzet:2020fd}, hot plates above the Leidenfrost temperature~\citep{Wachters:1966vy,Biance:2011di} and, more recently, cold plates covered with liquid nitrogen to exploit the cold Leidenfrost effect \citep{Antonini2013,Arora2018,Louhichi2020,Charles2021}. We have demonstrated elsewhere \citep{Louhichi2020} that, in the so-called viscous regime, biaxial extensional dissipation dominates the sheet dynamics produced using the cold Leidenfrost effect. Nature and pertinent industrial problems are more complex than the above two extreme cases. Hence, we may consider a third scenario, where drops impact a solid target of size comparable to that of the drop. In this case, a part of the expanding sheet is in contact with the target surface and the other part is expanding freely in the air, which, depending on the sample viscosity, may be the largest part of the sheet~\citep{Rozhkov2002,Rozhkov2003,Rozhkov2004,Rozhkov2006,
Villermaux2011,Vernay2015,Arora2016,Keshavarz:2016gb,Wang2018,Arogeti2019,Raux2020}. Although the situation of impacting drops on small targets was originally conceived to reduce the friction dissipation \citep{Rozhkov2002}, it turns out that it actually highlights the competition between shear dissipation (on the target) and biaxial extensional dissipation (in the air).
 In this work, we assess this competition by evaluating both dissipations (shear and biaxial extensional) for Newtonian and rheologically thinning non-Newtonian fluids. We show that, for Newtonian fluids, the shear and biaxial extensional dissipations are of the same order of magnitude. However, for the non-Newtonian thinning samples, the biaxial extensional dissipation is found to control the sheet expansion dynamics. We compare our data with experimental measurements of biaxial extensional viscosity of thinning fluids~\citep{Walker1996,Venerus2019,Louhichi2020} and model semi-quantitatively the sheet expansion dynamics using a biaxial extensional thinning viscosity.

\section{Materials and Methods}

\subsection{Materials}
\subsubsection{Newtonian fluids}
We investigate two classes of Newtonian fluids, silicone oils and mixtures of water and glycerol. Silicone oils, with zero-shear rate viscosities from $4.5$ mPa s to $339$ mPa s, an average surface tension of $20$ mN/m~\citep{Crisp1987} and densities ranging from 0.913 g/mL and 0.97 g/mL, are purchased from Sigma Aldrich and used as received. Glycerol/water mixtures with concentrations ranging from $22$ to $100$ \% g/g glycerol are used, yielding zero-shear rate viscosities from $1.7$ mPa s to $1910$ mPa s, densities from $1.05$ g/mL to $1.25$ g/mL~\citep{Association1963}, and an average surface tension of $65$ mN/m (as measured with a pendant drop set-up). 
	
\subsubsection{Non-Newtonian fluids}

\begin{figure}
	\centerline{\includegraphics[width=8.3cm]{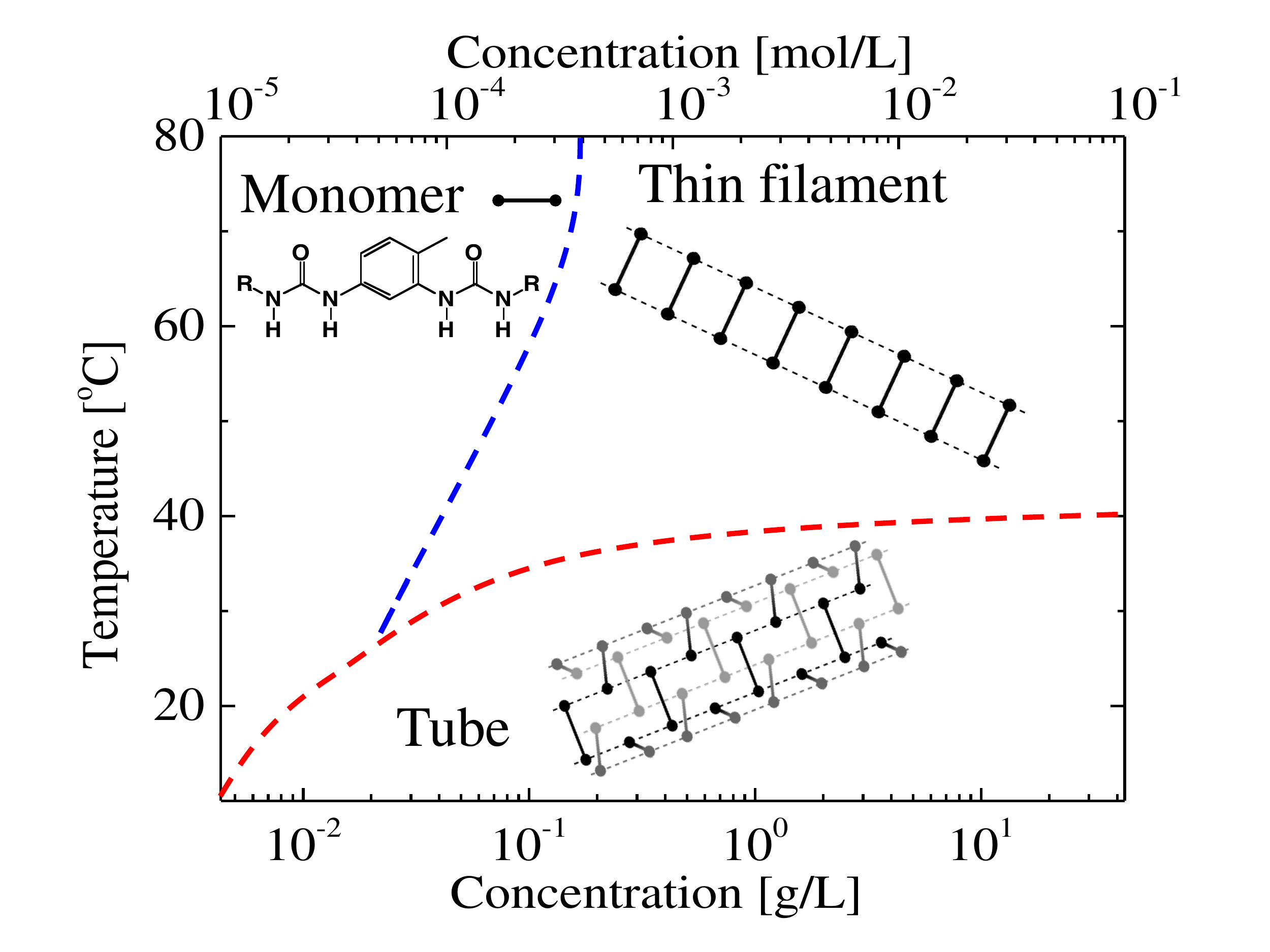}}
	\caption{ Phase diagram for EHUT solutions in toluene \citep{Bouteiller2005} showing the transition between monomers and supramolecular filaments. 
The structure of the EHUT monomer is shown, along with a schematic illustration of the various aggregates formed (out of scale), i.e., filaments and tubes. Hydrogen bonds are represented by dotted lines connecting the urea functions (black and gray circles). }
\label{figure:1}
	\end{figure}

	As non-Newtonian system, we choose wormlike micellar solutions (WLM) made of 2, 4-bis (2-ethylhexylureido) toluene, abbreviated as EHUT, dispersed in dodecane. This monomer has the ability to self-associate by means of hydrogen bonding in an apolar solvent (dodecane) forming supramolecular polymers~\citep{Isare2016,Lortie2002,Bouteiller2005,Ducouret2007,Shikata2008,Louhichi2017,Burger:js}. As schematically shown in Figure \ref{figure:1}, EHUT molecules self-assemble into small and long rod-like structure whose cross-section is solvent- and temperature-dependent. Above critical concentration and temperature, the monomers self-assemble into thin filaments with a diameter (1.3 nm) comparable to the size of one EHUT monomer. At lower temperatures, thick tubes with a cross-section equivalent to about three EHUT monomers form. In this work, we investigate samples in the tube region of the phase diagram, in the temperature range between 20$^\circ$C and 25$^\circ$C, and concentration range from $C$=$0.37$ g/L to $C$=$3$ g/L, where the tubes are long enough to entangle, allowing us, hence, to explore a wide  range of viscoelastic properties. Although, this phase diagram is made for EHUT supramolecular assemblies formed in toluene, Fourier transform infrared spectroscopy (FTIR), small angle neutron scattering (SANS) and rheology studies confirm that the range of temperatures and concentrations investigated here also correspond to the tube region of the EHUT assemblies formed in dodecane \citep{Bouteiller2005,Ducouret2007,Shikata2008}. The surface tension for EHUT solutions is assumed to be independent of concentration. We take $\Gamma=25 \pm 2$ mNm$^{-1}$ as measured with a pendant drop experiment for a dilute sample ($C$ = $0.37$ g/L). The density of the EHUT solutions is assumed to be equal to the density of the solvent, $\rho = $ 0.75 g/mL.
	
\subsection{Methods}
\subsubsection{Solution rheology}

Linear and nonlinear shear rheology measurements are performed with a MCR501 stress-controlled rheometer (Anton Paar, Austria), operating in the strain-control mode and equipped with a stainless steel cylindrical Couette geometry. Temperature control ($\pm 0.2^\circ $C) is achieved by means of a Peltier element.

For Newtonian and non-Newtonian samples, the zero-shear rate viscosity, $\eta_0 $, is measured by applying a ramp of steady shear rate varying from $0.01$ s$^{-1}$ to $1000$ s$^{-1}$.

For EHUT solutions, the linear viscoelastic spectra are obtained by applying a small-amplitude sinusoidal strain ($\gamma= 10\%$) with varying angular frequency, $\omega$, from $100$ to $0.01$ rad/s, and measuring the storage, $G'(\omega)$, and loss, $G''(\omega)$, moduli. The complex viscosity is also calculated from the linear viscoelastic spectra as $ | \eta^\star (\omega)|=\frac{\sqrt{(G'^2 (\omega)+G"^2 (\omega) )}}{\omega} $.~All experiments are performed in open air without any particular humidity precaution (the relative humidity is about $40\%$).~Since hydrogen bonding systems such as EHUT self-assemblies are very sensitive to humidity \citep{Louhichi2017a,Van-Zee:2018aa}, and because we do not control the drop impact environmental conditions, we ensure that both bulk rheology and drop impact experiments are conducted under the same temperature and humidity conditions.


\subsubsection{Drop impact experiment} 

\begin{figure}
  \centerline{\includegraphics[width=8.3cm]{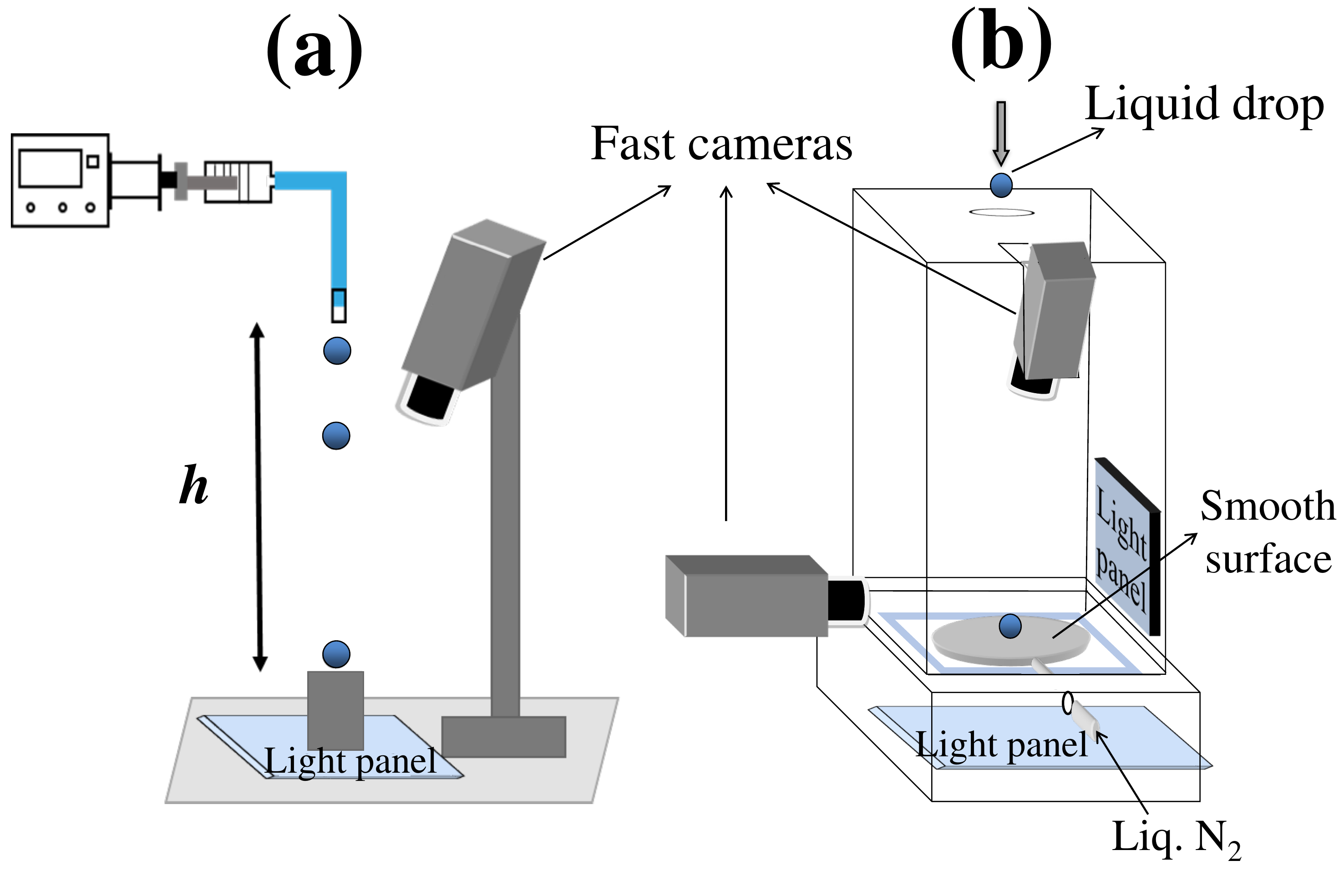}}
\caption{a) Schematic illustration of the main experimental setup configuration, consisting of a drop impacting a small solid target as described in the text. b) Schematic illustration of the second experimental setup showing a drop falling on a liquid nitrogen thin layer. The expansion event is recorded using two fast cameras allowing concomitant top and side visualizations.}
\label{figure:2}
\end{figure}

To study the competition between shear and biaxial extensional viscous dissipations, two drop impact experimental set-ups are used. The main set-up based on the impact on a small cylindrical target was originally designed by Vignes-Adler et al. \citep{Rozhkov2002} and subsequently modified by several groups \citep{Villermaux2011, Vernay2015, Wang2017}. We use the same set-up as in \citep{Vernay2015, Arora2016}, schematically shown in Figure \ref{figure:2}a. In brief, a hydrophilic glass target of diameter $d_T=6.5$ mm is fixed on top of an aluminum rod with the same diameter. The liquid is injected from a syringe pump with a flow rate of $1$ mL/min through a needle placed vertically above the target. The initial diameter of the falling drop is constant, $d_0=3.9$ $\pm$ $0.2$ mm, as measured by image analysis and by weighting the drops. Note that, considering the difference in fluids' surface tensions, we have adapted the needle diameter to produce a constant drop diameter (A needle with a diameter of 2 mm for fluids with a surface tension of $65$ mN/m and a needle diameter of 4 mm for fluids with $\Gamma=20$ - $25$ mN/m). The drop falls from a height $h=91$ cm, yielding an impact velocity $v_0=\sqrt{2gh}=4.2 $ ms$^{-1}$ ($g$ is the acceleration of gravity). The corresponding Weber numbers, $\textrm{We}=\frac{v_0^2\rho d_0}{\Gamma}$, are We $\sim1200$ for glycerol/water mixtures, We $\sim3400$ for silicone oils and We $\sim2400$ for EHUT-based samples. The target is mounted on a transparent Plexiglas plate, illuminated from below by a high-luminosity backlight (Phlox LLUB, luminance of 20 cd/m$^2$). The drop impact is recorded from the top using a high-speed camera (Phantom V 7:3) operating at $6700$ frames/s with a resolution of $800 \times 600$ pixels$^2$. The angle between the camera axis and the horizontal plane is fixed at about $10^{\circ}$.

Additional experiments are performed with Newtonian fluids under cold Leidenfrost conditions \citep{Antonini2013}. The relevant set-up is described elsewhere \citep{Louhichi2020} and schematized in Figure \ref{figure:2}b. In brief, the drop maintained at room temperature impacts a smooth surface (polished silicon wafer or quartz slide with a diameter larger than the maximum sheet diameter) covered with a thin layer of liquid nitrogen (temperature $T$ = $-196$ $^\circ $C). Upon impact a vapor cushion forms at the liquid interface due to the evaporation of N$_2$, providing a unique scenario of non-wetting and slip conditions that eliminates shear viscous dissipation \citep{Antonini2013,Chen2016}. Details on the heat transfer involved in the impact of drops in cold Leidenfrost conditions has been published elsewhere \citep{Charles2021}, it does not change significantly the temperature of the drops during the expansion phase of the sheet.

\subsubsection{Image analysis}
The time evolution of the diameter of the sheet is measured with ImageJ software. We first subtract the background image from the expansion movie and highlight the rim by a binary thresholding. This allows us to determine the contour of the sheet and measure its area $A$. An apparent sheet diameter is simply deduced as: $d=\sqrt{\frac{4A}{\pi}}$. The results are  obtained by averaging for each sample the time evolution of the sheet diameter from three different experiments. Note that for the small target setup, the dark black central disk is the target an precludes to study the expansion of sheets that do not  cross the edge of the target.  For the cold Leidenfrost set-up, in addition to the top view images, systematic side view images are also collected, though not shown here, to correct the maximum expansion diameter for the corona effect observed for the low viscosity samples (viscosity smaller than 100 mPa s), as detailed in \citep{Louhichi2020}.

\section{Results and discussions}

\subsection{Newtonian fluids}

\subsubsection{Drop impact experiments}

\begin{figure}

\centerline{\includegraphics[width=8.3cm]{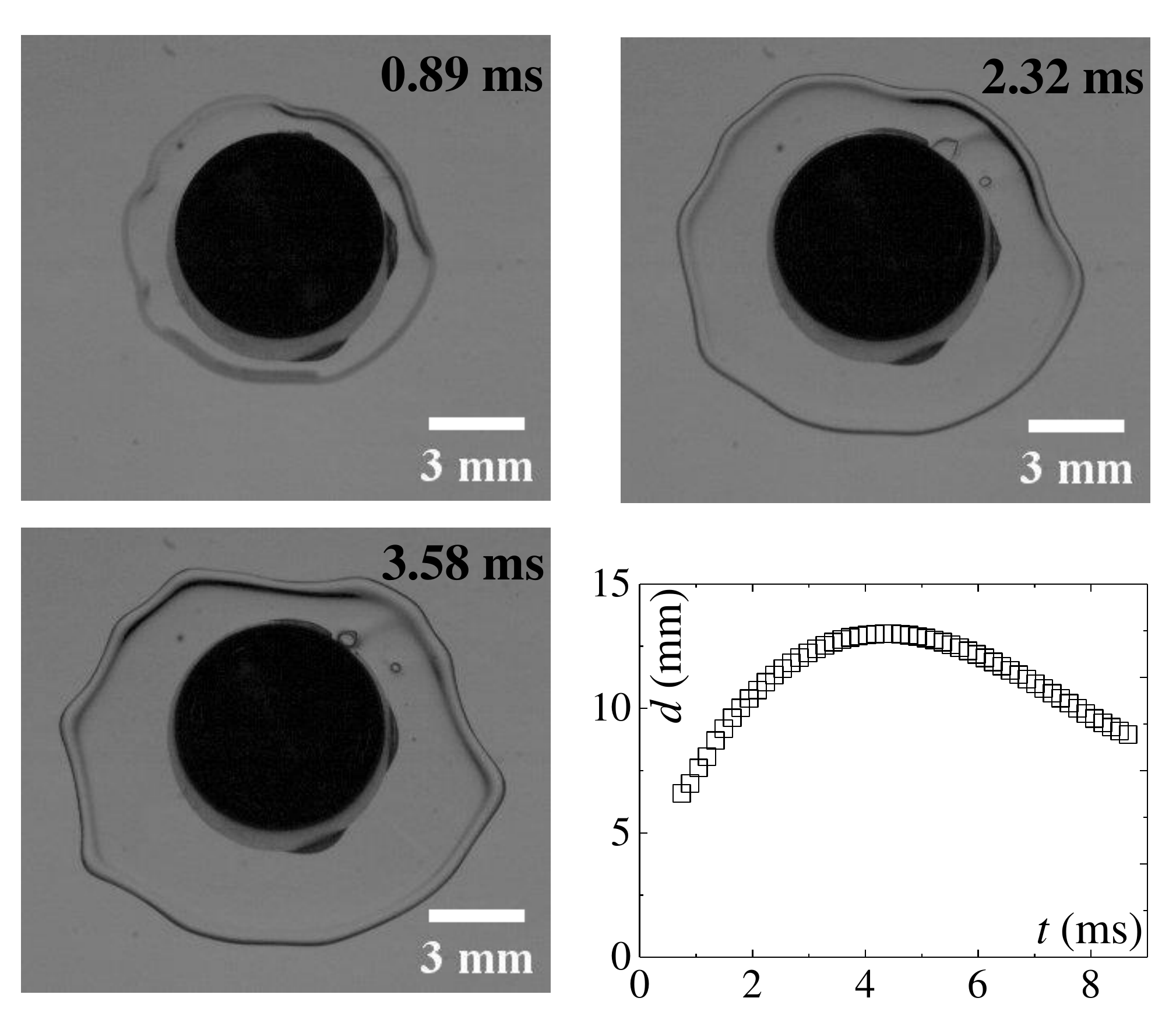}}

\caption{Snapshots taken at different times (as indicated on each image) during the expansion of the sheet for a glycerol/water mixture with a viscosity $\eta_0 = 160$ mPa s produced upon impact on a solid target with a diameter $d_T=6.5$ mm. The maximum expansion occurs at $t_{\rm{Max}} = 3.58$ ms.The black area represents the surface of the target. (Multimedia view)}
\label{figure:3}
\end{figure}

\begin{figure}

\centerline{\includegraphics[width=8.3cm]{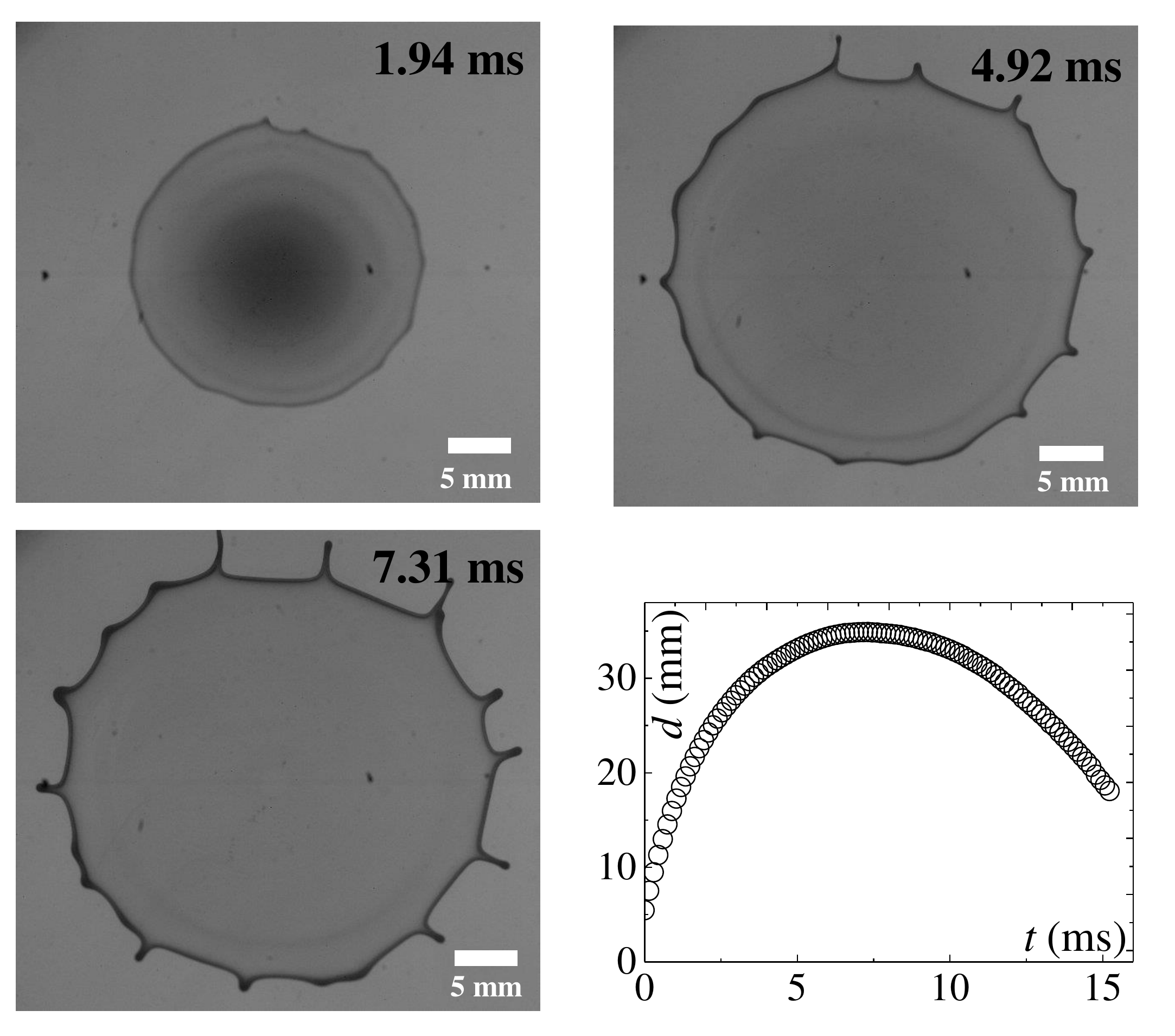}}

\caption{Snapshots taken at different times (as indicated on each image) during the expansion of the sheet for a glycerol/water mixture with a viscosity $\eta_0 = 160$ mPa s  produced upon impact under cold Leidenfrost conditions. The maximum expansion occurs at $t_{\rm{Max}}= 7.31$ ms. (Multimedia view)}

\label{figure:4}
\end{figure}

Once hitting the target, the drop spreads first on the target and then expands freely in the air until reaching a maximum expansion. It then retracts because of surface tension and eventually bulk elasticity for non Newtonian fluids (representative videos of the impact process are available in the supplementary materials). This general phenomenology has been observed for many types of fluids impacting a small solid target \citep{Rozhkov2002,Villermaux2011,Luu2013,Vernay2015,Arora2016,Wang2017,Wang2018}. The overall behaviour is illustrated in Figure \ref{figure:3} (Multimedia view) that depicts snapshots of the drop after its impact for a Newtonian sample (glycerol/water) with $\eta_0 = 160\pm10$ mPa s at different times. 
In the bottom right of Figure \ref{figure:3} (Multimedia view), we show the corresponding time evolution of the effective sheet diameter, $ d $, where the expansion and retraction regimes can be clearly determined (we choose for the origin of time, the time when the drop hits the target).
Similarly, in Figure \ref{figure:4} (Multimedia view), we present snapshots for the same Newtonian sample at different times, when it expands and retracts under cold Leidenfrost conditions obtained by impacting a drop on a thin layer of liquid N$_2$. The origin of time is set when the droplet start expanding on the Nitrogen vapour cushion.

Despite qualitatively similar behaviour, we find that, for a same sample and a same impact velocity, the maximum expansion diameter of the sheet, $d_{\rm{Max}}$, is significantly larger under cold Leidenfrost conditions ($d_{\rm{Max}}=34.6$ mm) than with the small target ($d_{\rm{Max}}=12.9$ mm). When the expansion results from the impact on the target, only the inner  portion of the sheet is at intimate contact with the target surface while its outer part is free in the air, resulting of a combination of two potential sources of viscous dissipation: shear on the target and biaxial extensional everywhere but predominantly in air. The situation is less complex when the sheet is produced under cold Leidenfrost conditions because the sheet is expanding on the liquid Nitrogen vapour, ensuring biaxial extensional deformation as the unique source of viscous dissipation\citep{Louhichi2020}.

The cold Leidenfrost conditions allow also the observation of the sheet central region, usually obscured by the target. It is worth to note the occurrence of a thicker rim for the sheet when impacted on the target and, on the other hand, the appearance of a larger  number of better defined fingers instabilities under cold Leidenfrost conditions. However, we will not discuss further these two interesting features, which are out of the scope of this work. Moreover, in the rest of the paper, we will focus on the sheets expansion dynamics up to their maximum expansion and not consider the retraction regime.

In order to follow the expansion dynamics of different samples, we focus on the maximum expansion diameter $d_{\rm{Max}}$. We use a normalized maximum expansion diameter, $ \tilde{d} $, adopting the same definition as in \citep{Arora2016,Louhichi2020}:

\begin{equation}
\label{eqn:dtilde}
\tilde{d}=\frac{d_{\rm{Max}}}{d^{\rm{cap}}_{\rm{Max}}}
\end{equation}	

\begin{figure}
\centerline{\includegraphics[width=8.3cm]{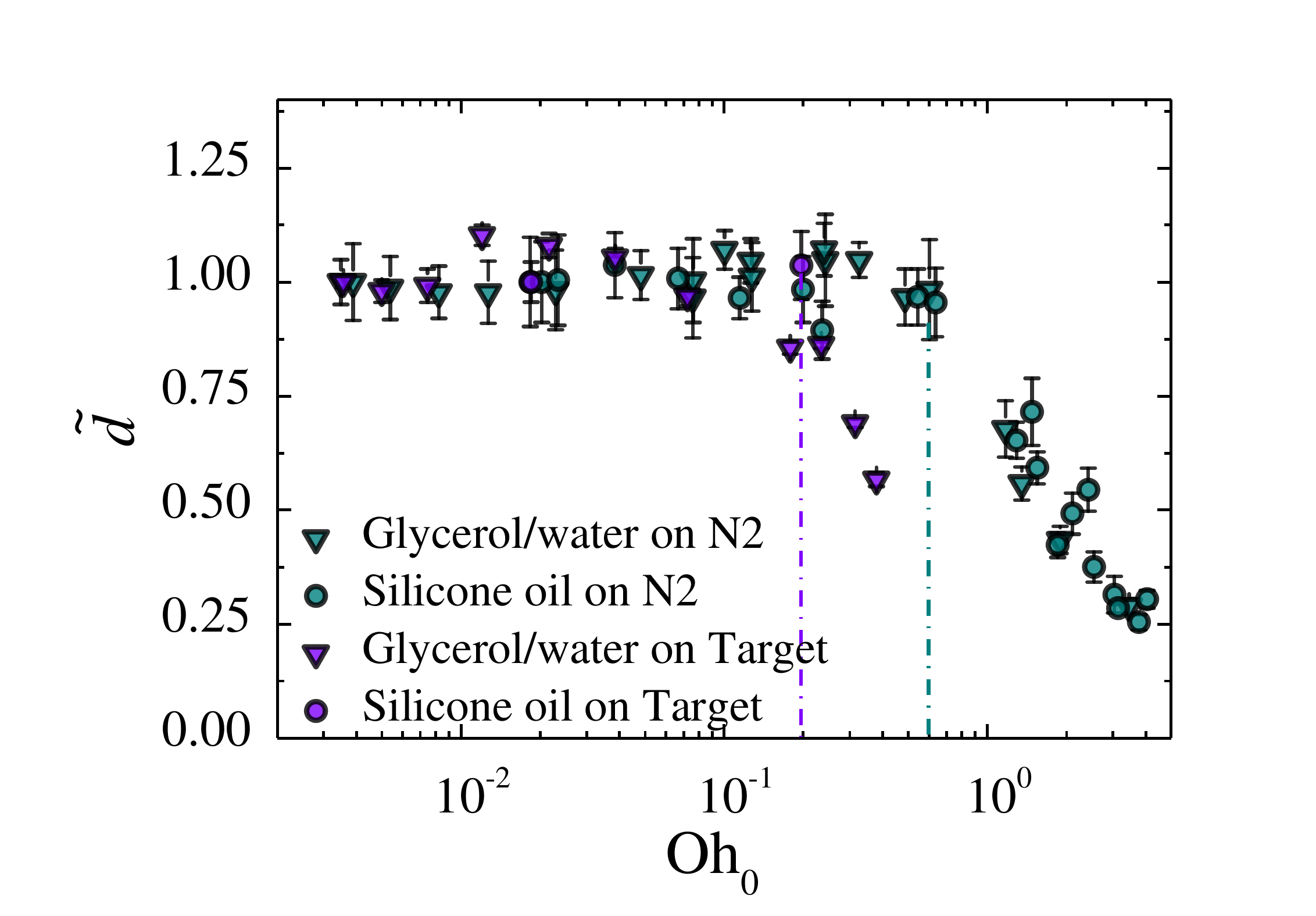}}

\caption{Normalized maximum expansion as a function of the zero-shear rate Ohnesorge number, Oh$_0$, for the two classes of Newtonian samples, as indicated in the legend, and for the two experimental configurations: small solid target (purple symbols) and cold Leidenfrost conditions (green symbols). Dotted lines indicate the onset of the viscous regime for each configuration. For the Newtonian samples (glycerol/water) presented in Figure \ref{figure:3} (Multimedia view) and Figure \ref{figure:4} (Multimedia view), Oh$_0=0.31$.}

\label{figure:5}
\end{figure}

Here $d_{\rm{Max}}$ is the maximum diameter of the sheet and $ d^{\rm{cap}}_{\rm{Max}} $ is the maximum expansion diameter in the capillary regime (at low viscosity), where viscous dissipation does not reduce significantly the maximum expansion diameter. This normalized quantity allows us to compare drops with different initial diameters $d_0$ and different surface tensions. In addition, to account for the variation of surface tensions among samples, the data are plotted against the dimensionless zero-shear rate Ohnesorge number, which expresses the ratio of viscous forces to inertial and surface tension forces: Oh$_0 =\frac{\eta_0}{\sqrt{\rho\Gamma d_0}}$, with $\rho $ the sample density, $\Gamma$ the surface tension and $\eta_0$ the zero-shear rate viscosity. Results for different Newtonian fluids obtained using two experimental conditions (target and cold Leidenfrost) are presented in Figure \ref{figure:5}. Note that, for the more viscous Glycerol/water mixtures ($\eta_0>200$mPa s), experiments on the solid target are not shown  because, the maximum expansion diameter is smaller than the target diameter itself. 
We find two regimes for both experimental configurations. The capillary regime where the dynamics is independent of viscosity resulting in a constant $\tilde{d}\approx1$, and the viscous regime, where  $\tilde{d}$  decreases with increasing viscosity. Importantly, the transition from capillary to viscous regime occurs at larger Oh$_0$ number (higher viscosity) for the sheets produced under cold Leidenfrost conditions (Oh$_0\approx0.6$) compared to those produced by impact on a target (Oh$_0\approx0.2$).
This implies that viscous dissipation is more important with the target setup than with the cold Leidenfrost setup, as exemplified by comparing the expansion dynamics of the Newtonian sample presented in Figure \ref{figure:3} (Multimedia view) and Figure \ref{figure:4} (Multimedia view).

In addition, Figure \ref{figure:5} confirms the fact that even in the absence of friction on a solid surface, a dissipation regime exists resulting in the decrease of $\tilde{d}$ with Oh$_0$. In this configuration, the deformation of the sheet is biaxial extensional \citep{Louhichi2020}. The dissipation on a small solid target is by contrast more complex because it results from the contact with a solid surface and from free expansion in the air (similar to that under cold Leidenfrost conditions). Thus, in the following section, we will analyse the contribution of each dissipation mechanism to the expansion dynamics of a Newtonian drop impacting a small solid surface.

 \subsubsection{Viscous dissipation processes}

\paragraph{Shear dissipation energy}

To estimate the shear dissipation energy, we first define the shear rate experienced by the sheet during its expansion on the small solid target. We approximate the expanding sheet by a disk. The shear flow velocity field on the target \citep{Madejski1976,Attane:2007fy} reads 

\begin{equation}
\left\{\begin{matrix}
v_r=\frac{2}{RH}\frac{dR}{dt}rz &(\textrm{radial}) \\ 
& \\
v_z=\frac{-2}{RH}\frac{dR}{dt}z^2 & (\textrm{axial})
\end{matrix}\right.
\end{equation}
 where $R(t)=\frac{d(t)}{2}$ is the radius of the expanding sheet, $r$ denotes the radial direction, $z$ denotes the axial direction and $H(t) =\frac{d_0^3}{6R(t) ^2}$ is the mean thickness of the expanding sheet in the disk shape approximation with $d_0$ the drop diameter.
 The instantaneous shear rate reads
 \begin{equation}
\dot{\gamma}=\frac{1}{2}\left(\frac{\partial v_r}{\partial z}+\frac{\partial v_z}{\partial r}\right)=\frac{1}{RH}\frac{dR}{dt}r
\end{equation}

The viscous shear energy dissipated, $E_S$, during sheet expansion process can be written as:

\begin{equation}
\label{eqn:ES}
E_{S}\approx \int_{0}^{t_{\rm{Max}}} \int_{V}\sigma_{S}\dot{\gamma}dVdt
\end{equation}	

where $V$ denotes the sheared volume and $\sigma_{S}$ is the shear stress. During the expansion of the drop sheet, the viscous boundary layer thickness, $\delta(t) = \alpha \sqrt{\frac{\eta_\mathrm{S} t}{\rho}}$, ($\alpha $ is an unknown prefactor and $\eta_{S}=\eta_0$ is the shear viscosity) which quantifies the thickness of the sheet that is actually sheared \citep{Roisman2009, Arora2016, Clanet2002}. At short time, $\delta<H$, but the viscous boundary layer grows with time, while the film thickness, $H(t)$, decreases. Hence, at some time $\delta$ exceeds $H$ and the relevant thickness to be considered is $H$ and not $\delta$ anymore.

For simplicity, here, we consider the two limits for the viscous shear energy: (i) $E_\textrm{S,H}$, where the  relevant thickness for the sheared volume is $H(t)$, and (ii)  $E_{\textrm{S},\delta}$, where the relevant thickness for the sheared volume is $\delta(t)$. $E_\textrm{S,H}$ reads  

\begin{eqnarray}
E_\textrm{S,H}&&=\int_0^{t_\textrm{max}}dt\int_0^{d_\textrm{T}/2}\eta_\textrm{S}\dot{\gamma}^22\pi rH dr\nonumber\\
&&=\frac{3\pi \eta_\mathrm{S}d_\textrm{T}^4}{16d_0^3}\int_0^{t_\textrm{max}}\left(\frac{dR}{dt}\right)^2 dt
\end{eqnarray}

With the approximations $\int_0^{t_\textrm{max}}\left(\frac{dR}{dt}\right)^2 dt \approx t_\textrm{max}\left \langle \left(\frac{dR}{dt}\right)^2\right \rangle$ , $t_\textrm{max}\approx \frac{d_\textrm{max}}{v_0}$ and $d_\textrm{max}\gg d_0$, $E_\textrm{S,H}$ reads

\begin{equation}
E_\textrm{S,H}\approx \frac{3\pi \eta_\mathrm{S}d_\textrm{T}^4v_0}{64d_0^3}d_\textrm{max}
\label{eqn:ESH1}
\end{equation}
On the other hand, $E_{\textrm{S},\delta}$ reads 

\begin{eqnarray}
E_\mathrm{S,\delta}&&=\int_0^{t_\textrm{max}}dt\int_0^{d_\textrm{T}/2}\eta_\textrm{S}\dot{\gamma}^22\pi r\delta dr\nonumber\\
&&=\frac{9\pi\alpha\eta_\textrm{S}^{3/2}d_\textrm{T}^4}{8\sqrt{\rho}d_0^6}\int_0^{t_\textrm{max}}R^2\left(\frac{dR}{dt}\right)^2 t^{1/2} dt
\end{eqnarray}

With the approximations $\int_0^{t_\textrm{max}}R^2\left(\frac{dR}{dt}\right)^2t^{1/2}dt\approx t_\textrm{max}\left \langle \frac{1}{4} \left(\frac{dR^2}{dt}\right)^2\right \rangle \left \langle t^{1/2}\right \rangle$, $t_\textrm{max}\approx \frac{d_\textrm{max}}{v_0}$ and $d_\textrm{max}\gg d_0$, $E_{\textrm{S},\delta}$ reads  

\begin{equation}
E_\mathrm{S,\delta}\approx\frac{3\pi\alpha\eta_\textrm{S}^{3/2}d_\textrm{T}^4v_0^{1/2}}{256\sqrt{\rho}d_0^{5/2}}\left(\frac{d_\textrm{max}}{d_0}\right)^{7/2}
\label{eqn:ESdelta1}
\end{equation}

\paragraph{Biaxial extensional dissipation energy}

A typical example of a biaxial extensional flow is the compression of a sample sandwiched between two (continuously) lubricated surfaces \citep{Macosko1994}. The biaxial viscosity is defined as $\eta_B=\frac{\sigma_{rr}-\sigma_{zz}}{\dot{\varepsilon}}$, where $\dot{\varepsilon} $ is the strain rate, and $\sigma_{rr}$ and $\sigma_{zz}$ are the stress tensor components in cylindrical coordinates \citep{Macosko1994}. The stress tensor difference $\sigma_{rr}-\sigma_{zz}$ is commonly expressed as the biaxial stress $\sigma_B$. Hence, the biaxial viscosity reads $\eta_B= \frac{\sigma_B}{\dot{\varepsilon}}$. For a Newtonian fluid $\eta_\textrm{B}=6\eta_\textrm{S}$ \citep{Macosko1994}.

The biaxial extensional energy, $E_B$, dissipated during the expansion of the sheet can be written as \citep{Louhichi2020}:

\begin{eqnarray}
E_\textrm{B}&&=\int_0^{t_\textrm{max}}dt\int_V\sigma_\textrm{B}\dot{\varepsilon}dV\nonumber\\
&& \approx\eta_\textrm{B}\frac{\pi d_0^3}{6}\int_0^{t_\textrm{max}}\left(\frac{1}{d}\frac{\partial d}{\partial t}\right)^2 dt\nonumber\\ 
&&\approx \eta_\textrm{B}\frac{\pi d_0^3}{6}t_\textrm{max}\left\langle \left(\frac{\partial \ln(d)}{\partial t} \right)^2 \right\rangle\nonumber\\
&&=\eta_\textrm{B}\frac{\pi v_0d_0^3}{6}\frac{\ln^2\left(\frac{d_\textrm{max}}{d_0}\right)}{d_\textrm{max}}
\label{eqn:EB}
\end{eqnarray}

where $\dot{\varepsilon}=\frac{1}{d}\frac{\partial d}{\partial t}$ is the biaxial extensional strain rate.

\begin{figure}[h]
\centerline{\includegraphics[width=8.3cm]{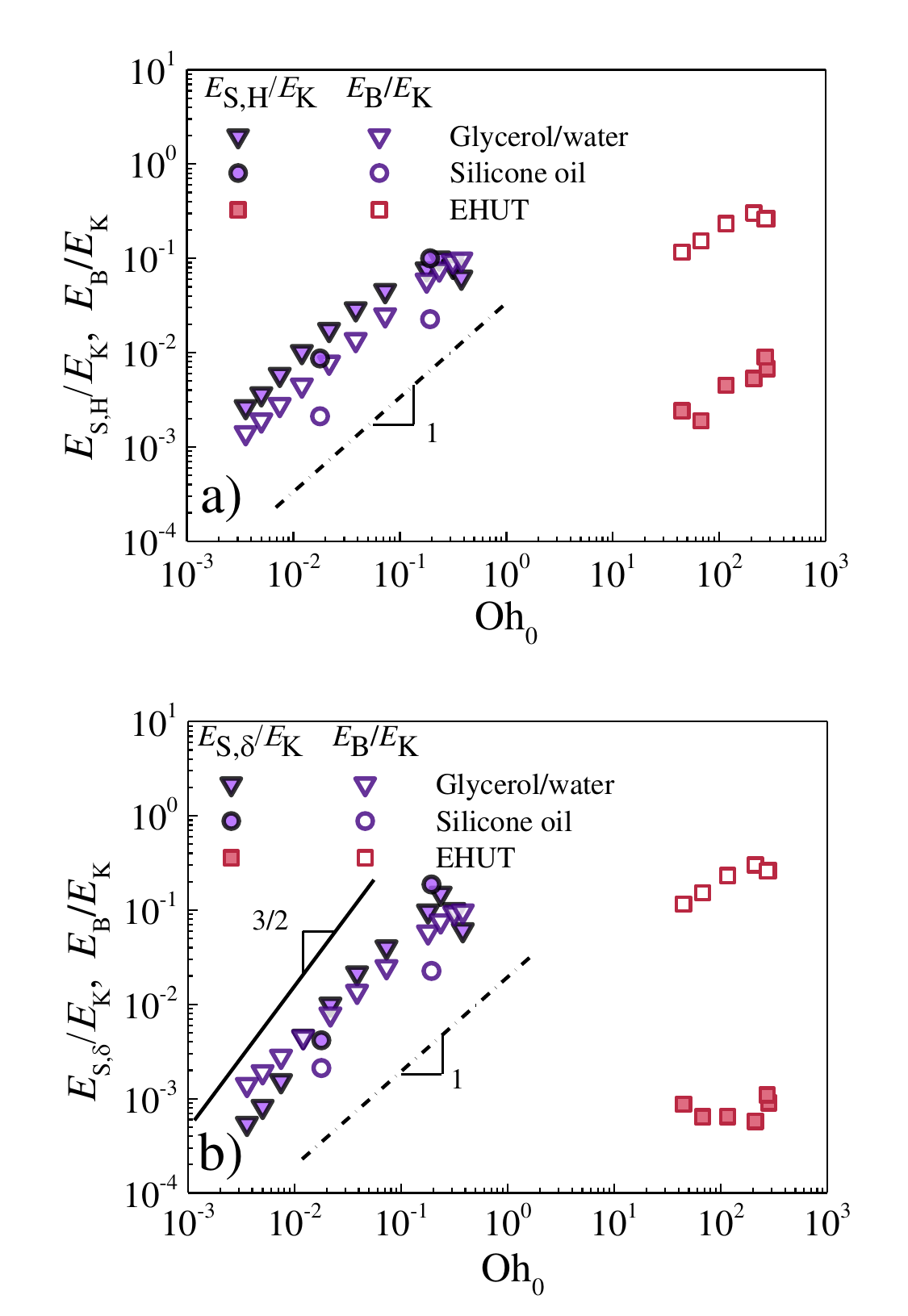}}
\caption{Variations of the biaxial extensional viscous energy $E_\textrm{B}$
 calculated  from Eq.(\ref{eqn:EB}) and  the two limits of the shear viscous dissipation energy respectively (a) $E_\textrm{S,H}$ from Eq. (\ref{eqn:ESH1}), and (b) $E_\mathrm{S,\delta}$ from Eq.(\ref{eqn:ESdelta1}) during the expansion of a drop impacting a small solid target, as a function of the zero-shear rate Ohnesorge number, Oh$_0$, for the two classes of Newtonian fluids (purple symbols) and EHUT supramolecular polymer solutions (red squares). The dissipation energies have been normalized by the initial kinetic energy of the drop $E_\mathrm{K}=\frac{1}{2}mv_0^2$.}The dashed and solid lines indicate a slope of 1 and 1.5, respectively.
\label{figure:6}
\end{figure}

\paragraph{Comparison between shear and biaxial extensional viscous dissipations}
In order to assess the respective roles of viscous dissipation due to shear and biaxial extensional deformations during the expansion of a drop, we compare in Figure \ref{figure:6} the variations  of the biaxial extensional viscous energy $E_\textrm{B}$ (Eq.(\ref{eqn:EB})) with the two limits for the shear viscous dissipation energy  $E_\textrm{S,H}$ (Eq.(\ref{eqn:ESH1})) (Figure \ref{figure:6}a) and $E_\mathrm{S,\delta}$ (Eq.(\ref{eqn:ESdelta1})) (Figure \ref{figure:6}b). The dissipation energies have been normalized by the initial kinetic energy of the drop $E_\mathrm{K}=\frac{1}{2}mv_0^2$, with $m$ the mass of the drop, and plotted as a function of the zero-shear rate Ohnesorge number, Oh$_0$. We find that $E_\textrm{S,H}$ and $E_\mathrm{S,\delta}$ exhibit power law evolution with the viscosity, which mirrors their explicit dependence with $ \eta_0$, $E_\textrm{S,H}\sim \eta_0$ (Eq.\ref{eqn:ESH1}) and $E_\mathrm{S,\delta}\sim \eta_0^{3/2}$ (Eq.\ref{eqn:ESdelta1}), reflecting the fact that the dependence of $d_\mathrm{Max}$ with viscosity is much weaker. This is also the case for $E_\textrm{B}$ which increases as a power law with $ \eta_0$, $E_\textrm{B} \sim \eta_0$ (Eq.\ref{eqn:EB}), since $ \eta_B$ = 6$ \eta_0$ for Newtonian fluids.

Moreover, $E_\mathrm{S,\delta}$ and $E_\textrm{S,H}$ are of the same order of magnitude. Hence, for convenience, we choose the simplest one, $E_\textrm{S,H}$ for further discussion on the competition between shear and biaxial extensional viscous dissipation. From Eqs (\ref{eqn:ESH1},\ref{eqn:EB}), we can compute the ratio of the two sources of viscous dissipation energy as 

\begin{equation}
\frac{E_\mathrm{B}}{E_\mathrm{S}}=\frac{64}{3}\left(\frac{d_0}{d_\mathrm{Max}}\right)^2\left(\frac{ d_0}{d_\mathrm{T}}\right)^4\ln^2\left(\frac{d_\textrm{max}}{d_0}\right)
\end{equation}

 This ratio strongly depends on the drop-to-target size ratio, $\frac{d_0}{d_\mathrm{T}}$. It does not depend explicitly on $\eta_0$, but indirectly because of the dependence of the maximal expansion diameter on the sample viscosity. Furthermore, as evidenced by the superposition of $E_\textrm{S,H}/E_\textrm{K}$ and $E_\textrm{B}/E_\textrm{K}$ in Figure \ref{figure:6}, $\frac{E_\textrm{B}}{E_\textrm{S,H}}$ is close to one and, as expected, depends only weakly on $\eta_0$.

\subsection{Viscoelastic supramolecular polymers}
In this section, we present the results for non-Newtonian fluids. In contrast to their Newtonian counterparts, these samples have shear and biaxial extensional viscosities that depend on the applied deformation rate.

\subsubsection{Viscoelasticity}
\begin{figure}
\centerline{\includegraphics[width=8.3cm]{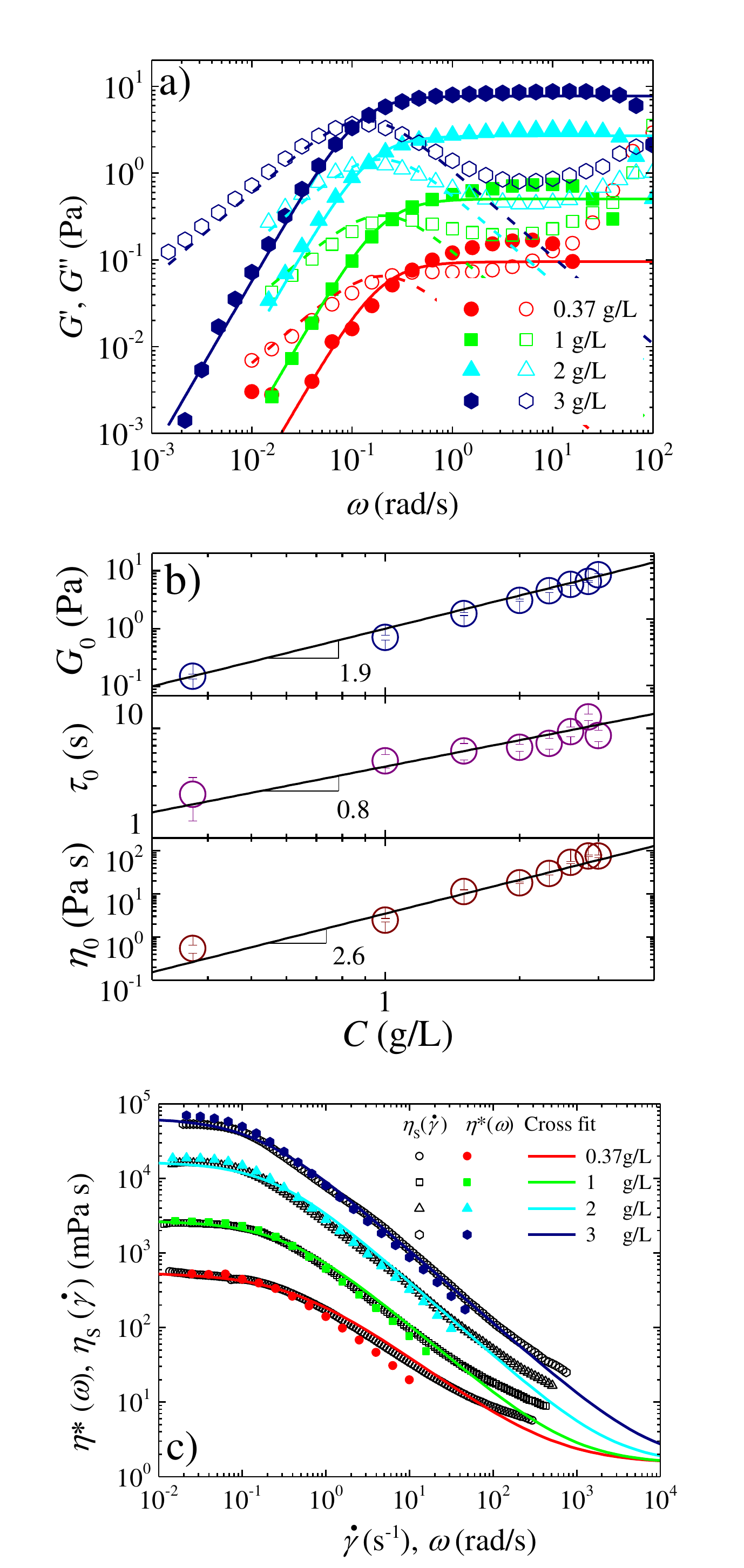}}
\caption{(a) Frequency dependence of the storage ($G'$, filled symbols) and loss ($G''$, open symbols) moduli of different EHUT supramolecular polymer solutions. The lines (solid for $G'$, dashed for $G''$) are Maxwell model fits for samples with different concentrations as indicated in the legend. (b) Evolution of the storage modulus (top), terminal relaxation time (middle), extracted from the Maxwell fits, and zero-shear rate viscosity (bottom) with sample concentration. The concentration  dependence is indicated by the slopes in each figure. (c) Complex viscosity as a function of frequency (closed symbols) and steady shear viscosity as a function of shear rate (open samples), and fits (lines) using the Cross equation (Eq.\ref{eqn:Cross}), for samples with different concentrations as indicated in the legend.}
\label{figure:7}
\end{figure}

Figure \ref{figure:7}a shows the dynamic moduli ($G'$, $G''$) as a function of oscillatory frequency ($\omega$) for some representative concentrations. As discussed previously \citep{Ducouret2007}, these systems behave as Maxwell fluids at low frequencies. This is confirmed by the good agreement between the data and the Maxwell fits (lines in Figure \ref{figure:7}a) where {$G' = G_0\frac{\omega^2\tau_0^2}{1+\omega^2\tau_0^2}$} and {$G'' = G_0\frac{\omega\tau_0}{1+\omega^2\tau_0^2}$}, with $G_0$ being the elastic modulus and $\tau_0$ the characteristic relaxation time. In Figure \ref{figure:7}b, we plot the relevant viscoelastic quantities ($G_0$, $\tau_0$ and $\eta_0$) for the range of concentrations studied here, extracted from the linear viscoelastic spectra. We find that these quantities follow scaling laws with concentration ($G_0 \sim C^{1.9}$, $\tau_0\sim C^{0.8}$ and $\eta_{0} \sim C^{2.6}$) in agreement with \citep{Louhichi2017}.

Figure \ref{figure:7}c depicts the complex viscosity, $\vert\eta^{*}\vert$, as a function of frequency, along with the steady shear viscosity, $\eta_S (\dot{\gamma})$, as a function of shear rate. The collapse of the dynamic and steady data (especially above 0.37 g/L) validates the Cox-Merz rule \citep{Cox1958}. We describe the data with a fit by means of the Cross model~\citep{Cross1965}, one of many options for empirical models, all giving comparable results:

\begin{equation}
\label{eqn:Cross}
\eta_S (\dot{\gamma})=\eta_{\infty} + \frac{\eta_0-\eta_{\infty}}{1+(k\dot{\gamma})^{n}}
\end{equation}

Here $\eta_{\infty}$ is the viscosity at very large shear rate that we set equal to the solvent viscosity ($\eta_{\infty} =1.5$ mPa s), $\eta_0$ is the zero-shear rate viscosity (plotted in Figure \ref{figure:7}b), $n$ is the shear-thinning exponent and the parameter $k$ is the inverse of a critical shear rate \citep{Macosko1994}. We find that the Cross model provides a good description of the samples' shear-thinning behaviour (continuous lines in Figure \ref{figure:7}c), except at very large shear rates where some deviation is systematically measured. All data can be fitted with very similar shear-thinning exponent $n=0.97\pm0.05$.

\subsubsection{Drop impact experiments}
\label{Drop impact experiments EHUT}

\begin{figure}[h!]

\centerline{\includegraphics[width=8.3cm]{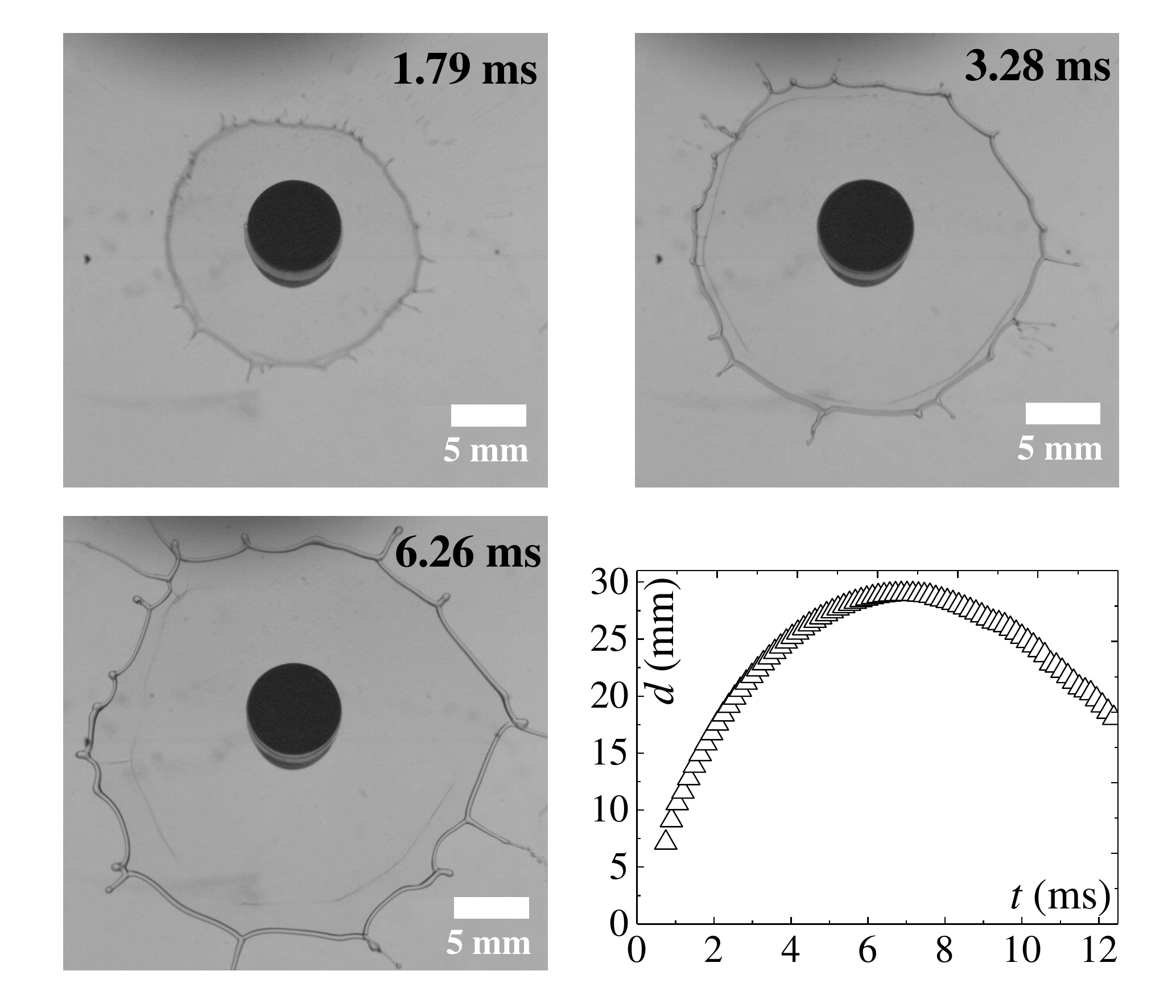}}

\caption{Snapshots taken at different times (as indicated on each image) during the expansion of a sheet for EHUT sample with concentration $C = 0.37$ g/L. (Multimedia view)}

\label{figure:8}
\end{figure}

Figure \ref{figure:8} (Multimedia view) depicts snapshots of the drop after its impact for a EHUT solution with $C = 0.37$ g/L at different times. The general phenomenology that has been observed for Newtonian samples (see Figure \ref{figure:3} (Multimedia view) and Figure \ref{figure:4} (Multimedia view)) is also found here, and the expansion process does not exhibit any pertinent qualitative differences compared to that of Newtonian samples. This suggests common physical processes driving the expansion dynamics of sheets of different kinds of systems.

\begin{figure}

\centerline{\includegraphics[width=8.3cm]{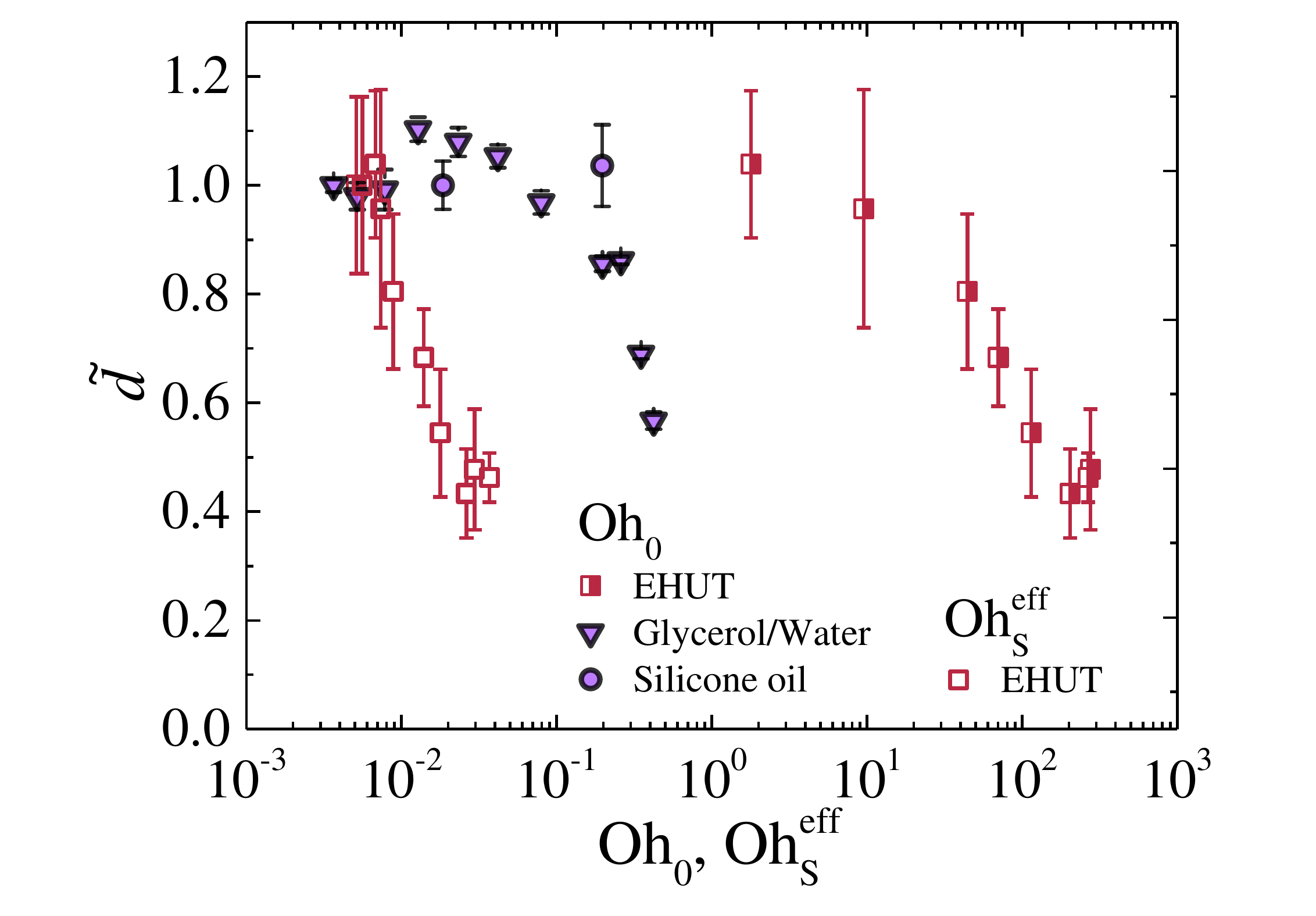}}

\caption{Normalized maximum expansion factor $\tilde{d}$ as function of the zero-shear rate Ohnesorge number, Oh$_0$, for a EHUT supramolecular polymer solutions (closed squares) and for the Newtonian water/glycerol mixtures and silicone oils (purple symbols) and of the  effective  shear  Ohnesorge number, Oh$_\mathrm{S}^{\mathrm{eff} }$, (see text) for the EHUT supramolecular polymer solutions (open squares).}
\label{figure:9}
\end{figure}


In Figure \ref{figure:9}, $\tilde{d}$ is plotted as a function of the zero-shear rate Ohnsesorge number, Oh$_0$ for both EHUT solutions (red half plain squares) and Newtonian samples (green plain triangles) impacted on a small target. We find that for the thinning fluids, the onset of the viscous regime takes place at much higher Oh$_0$, hence at much larger {$\eta_0$} (by factor of $100$) as compared to Newtonian samples, a clear indication that the zero-shear rate viscosity, is not the relevant quantity to characterize the dissipation process for viscoelastic fluids.
We first argue that elasticity does not play an important role in the expansion dynamics of the present viscoelastic sheets. Indeed, even if the characteristic relaxation time of the EHUT is always much larger than the characteristic time of the impact experiment (See the regime where storage modulus overwhelms the loss modulus in Figure \ref{figure:7}a) so that in principle elasticity should play a role in drop impact experiments, bulk elastic energy contribution is negligible with respect to the capillary contribution if the elastocapillary length, defined as $l_{ec}=\frac{3\Gamma}{G_0}$ is larger than the drop diameter $d_0$~\citep{Arora2018}. Taking $G_0=8.5 $ Pa, $\Gamma$=25 mN/m, and $d_0$=4 mm, for the most concentrated sample investigated (Figure \ref{figure:7}a,b), one gets $l_{ec}/d_0=2.2>1$. This ratio is even larger for less concentrated polymer solutions. Therefore, bulk elastic energy contributions can be safely neglected in the present framework. Hence, the contrasted results for Newtonian and non-Newtonian samples must stem from the thinning character of EHUT solutions. Actually, different groups have successfully accounted for the shear-thinning upon impact on a solid surface with a size larger than the expanding maximum sheet diameter~\citep{German2009,An2012,Andrade2015,Arora2016}.

In our case, the non-stationary shear rate experienced by the part of the expanding liquid sheet in contact with the target can be analyzed as following. We estimate the spatial average shear rate $\overline{ \dot{\gamma} (t)}$ experienced by the sheet on the target at a given time $t$. We assume a constant thickness $H(t)$ throughout $r$ at a given time. 
We define by $t^{\ast}$, the time at which the expanding sheet reaches the edge of the target, i.e. $R(t^{\ast})=R_\textrm{T}= \frac{d_\textrm{T}}{2}$.

 For $t<t^*$
\begin{equation}
\overline{ \dot{\gamma} (t)}=\frac{1}{\pi R^2(t)}\int_0^{R(t)}\frac{1}{R (t)H(t)}\frac{dR}{dt}2\pi r^2dr=\frac{4R^2(t)}{d_0^3}\frac{dR}{dt}
\end{equation}

 For $t>t^*$
\begin{equation}
\overline{ \dot{\gamma} (t)}=\frac{1}{\pi R_\textrm{T}^2}\int_0^{d_\textrm{T}/2}\frac{1}{R(t)H(t)}\frac{dR}{dt}2\pi r^2dr=\frac{2d_\textrm{T}R(t)}{d_0^3}\frac{dR}{dt}
\end{equation}
Finally, we calculate the time averaged shear rate experienced by the sheet on the target during the entire expansion process as

\begin{equation}
\left \langle \bar{ \dot{\gamma}} \right \rangle=\frac{1}{t_\mathrm{max}}\left(\int_0^{t^*}\frac{4R^2}{d_0^3}\frac{dR}{dt}dt+\int_{t^*}^{t_\mathrm{max}}\frac{4R_\textrm{T}R}{d_0^3}\frac{dR}{dt}dt\right)
\end{equation}

\begin{equation}
\left \langle \bar{ \dot{\gamma}} \right \rangle=\frac{4}{t_\textrm{max}d_0^3}\left(\frac{d_\textrm{T}^3}{24}+\frac{d_\textrm{T}\left(d_\textrm{max}^2-d_\textrm{T}^2\right)}{16}\right)
\end{equation}

\begin{figure}
	  \centerline{\includegraphics[width=8.3cm]{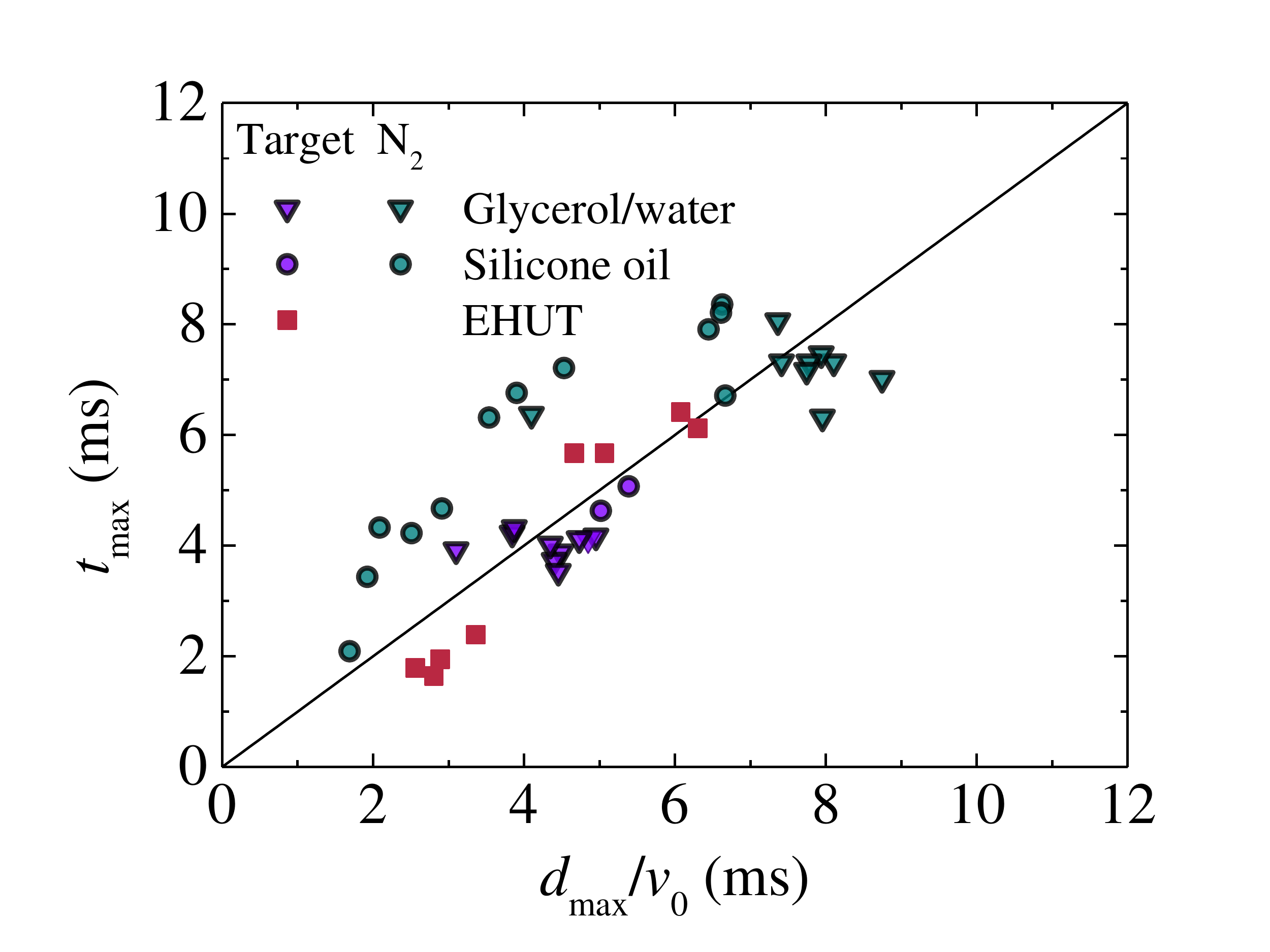}}
	\caption{Time to reach the maximal expansion, $t_\textrm{max}$, as a function of $d_\textrm{max}/v_0$, with $d_\textrm{max}$ the maximal expansion of the sheet and $v_0$ the impact velocity, for all impact experiments. The line corresponds to $t_\textrm{max}=\frac{d_\textrm{max}}{v_0}$.}
	\label{figure:tmax}
	\end{figure}

A good approximation, well supported by the experimental results (see Figure \ref{figure:tmax}) is $t_\textrm{max}\approx \frac{d_\textrm{max}}{v_0}$ with $v_0$ the impact velocity. Moreover, by assuming that $3d_\textrm{max}^2>>d_\textrm{T}^2$, the time averaged shear rate, $\left \langle \bar{ \dot{\gamma}} \right \rangle$, reads:

\begin{equation}
\left \langle \bar{ \dot{\gamma}} \right \rangle\approx\frac{d_\textrm{T}d_\textrm{max}v_0}{4d_0^3}
\label{eqn:gammamean}
\end{equation}

Having shown in Figure \ref{figure:7}c that the flow curves are well fitted by the Cross model (Eq.\ref{eqn:Cross}), the effective shear viscosities, $\eta_\mathrm{S}^\mathrm{eff}$, are evaluated from the Cross fits at the relevant shear rate,  $ \left \langle \bar{ \dot{\gamma}} \right \rangle $. The latter vary from  $ \left \langle \bar{ \dot{\gamma}} \right \rangle $ = 1420 s$^{-1}$ to 3220 s$^{-1}$ leading to effective viscosities varying between $\eta_\mathrm{S}^\mathrm{eff}=$ 10 mPa s and 2 mPa s.

In Figure \ref{figure:9}, we plot the normalised maximum expansion as a function of an effective Ohnesorge number, Oh$_\mathrm{S}^{\mathrm{eff} }$, where $\eta_0$ is replaced by $\eta_\mathrm{S}^\mathrm{eff}$ for the non-Newtonian fluids (open red squares). The plot shows that viscoelastic samples expand much less than one would expect based only on their shear viscosity at the relevant shear rate, when taking as reference the data for Newtonian fluids. Moreover, taking into account the shear-thinning viscosity generates an unrealistic situation where samples with almost the same shear-thinning viscosity exhibit different expansions. Therefore, the shear-thinning viscosity fails to properly describe the maximum expansion of the viscoelastic sheets. Instead the biaxial extensional viscous dissipation during the expansion in the air has to be considered. The deformation field during sheet expansion upon impacting a small target is essentially biaxial extensional, except on the small target where the sheet experiences both shear and biaxial extensional deformations. Given the very low value of the shear thinning viscosity, we anticipate that the dissipation is mainly due to the biaxial extensional flow.

\subsubsection{Viscous dissipation processes}

In order to estimate the shear and biaxial extensional dissipation energy of rheo-thinning fluids, one must asses the relevant rate dependent viscosities involved in these processes. \\

\paragraph{Effective shear viscosity}

In order to account for the rate-dependent viscosity in the shear dissipation energy of non-Newtonian fluids, we will consider the effective shear viscosity, $\eta_\textrm{S}^\textrm{eff}$, resulting from the shear thinning of the EHUT solution and accessible by mean of classical rotational rheometry (see section \ref{Drop impact experiments EHUT}).
   
\paragraph{Effective biaxial extensional viscosity}

\begin{figure}

\centerline{\includegraphics[width=8.3 cm]{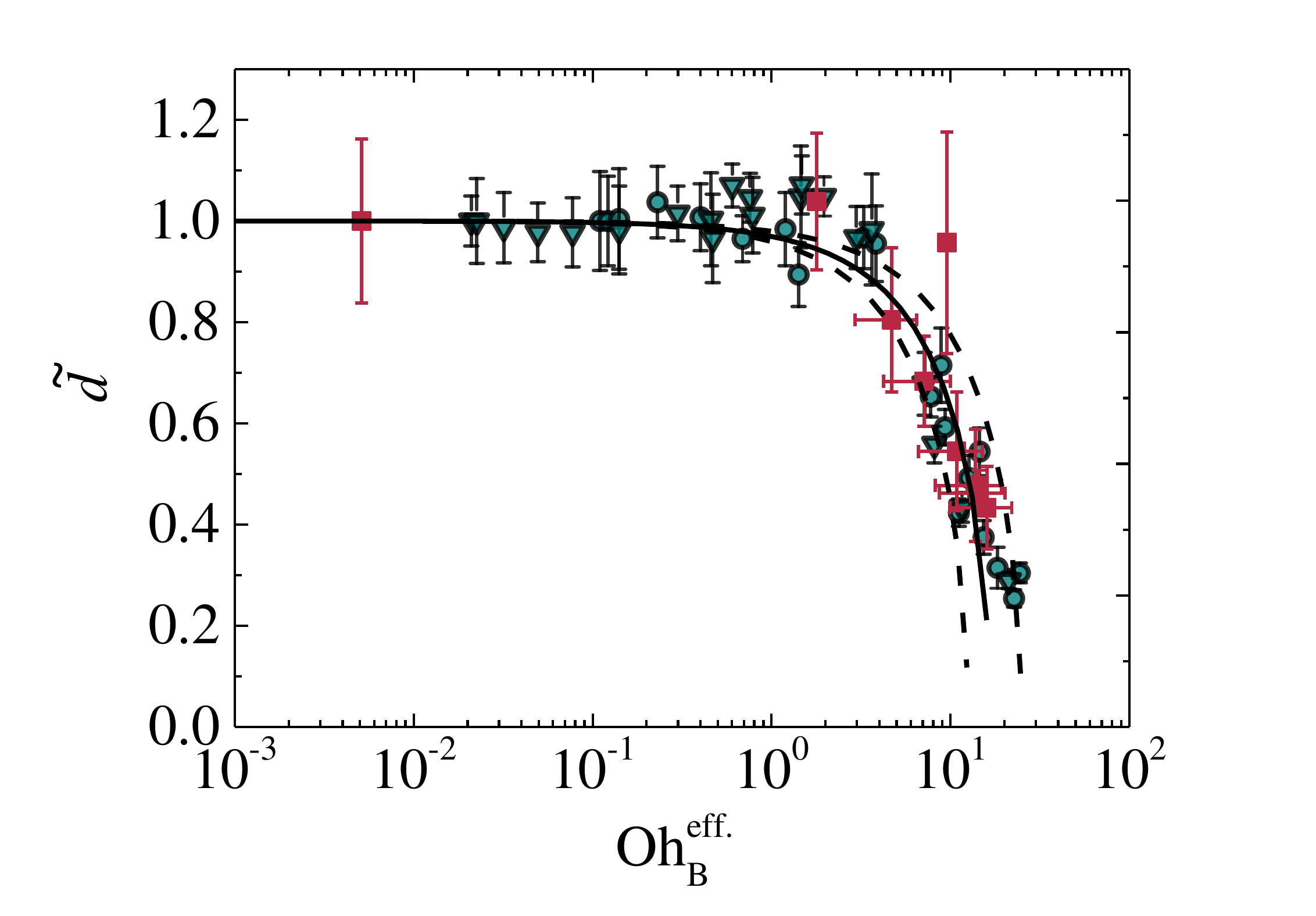}}

\caption{Normalized maximum expansion factor for drops of Newtonian fluids impacting a surface covered with liquid nitrogen as a function of Ohnesorge number based on $\eta^{B}_{0}$ (green symbols), together with that of EHUT samples (red squares) impacting a small target as a function of an effective Ohnesorge number based on an effective biaxial extensional viscosity $\eta\mathrm{^{eff}_{B}}$ chosen to collapse on the two sets of data in the viscous regime. The thin continuous line is the best fit with Eq.\ref{eqn:dtilde2}, and the dash lines show error bars on the fitting parameter $\beta$.}
\label{figure:11}
\end{figure}

Measuring properly the biaxial extensional viscosity is a difficult task for viscoelastic fluids of low viscosity such as the present EHUT solutions \citep{Walker1996,Joye1972,Venerus2010,Venerus2019,Johnson2016,Huang1993,Hachmann2003,Cathey1988,Maerker1974}.
In the framework of the impact drop problem, we have very recently established a successful, yet simple strategy in order to address this challenge \citep{Louhichi2020}. Our approach consists of mapping the viscoelastic systems to Newtonian samples impacted under cold Leidenfrost conditions where they experience a purely biaxial extensional deformation. Thus, we attribute to the viscoelastic sample an effective biaxial extensional viscosity equal to the biaxial extensional viscosity of a Newtonian fluid with the same maximum expansion factor. Operationally, we shift horizontally the data of $\tilde{d}$ obtained for the viscoelastic samples so that they overlap the reference data for the Newtonian fluids (Figure \ref{figure:11}). The shift factor yields directly an effective biaxial extensional Ohnesorge number, Oh$\mathrm{^{eff}_B}=\frac{\eta\mathrm{^{eff}_B}}{\sqrt{\rho\Gamma d_0}}$, from which an effective biaxial extensional viscosity $\eta\mathrm{^{eff}_B}$ is derived. $\eta\mathrm{^{eff}_B}$ is different from the biaxial extensional viscosity in the Newtonian limit, $\eta^0_{\mathrm{B}} = 6\eta_0$. Note that only $\tilde{d}$ data in the viscous regime are shifted for EHUT systems (data in the capillary regime are not subjected to the shift).

In Figure \ref{figure:11}, we show the fit of the master curve, with the following functional form, already established in \citep{Louhichi2020}:

\begin{equation}
\label{eqn:dtilde2}
\tilde{d}=\sqrt{1-\beta \mathrm{Oh^{eff}_B}}
\end{equation}

Here, $\beta=0.060\pm0.020$ is the only fitting parameter. The model in Eq.\ref{eqn:dtilde2} is based on an energy conservation balance assuming that the initial kinetic energy, $E_\textrm{K}$, is fully converted into surface energy, $E_{\Gamma}$, and biaxial extensional viscous dissipation energy, $E_\textrm{B}$ (the shear viscous dissipation energy, $E_\textrm{S}$, is neglected). 

\begin{figure}

\centerline{\includegraphics[width=8.3 cm]{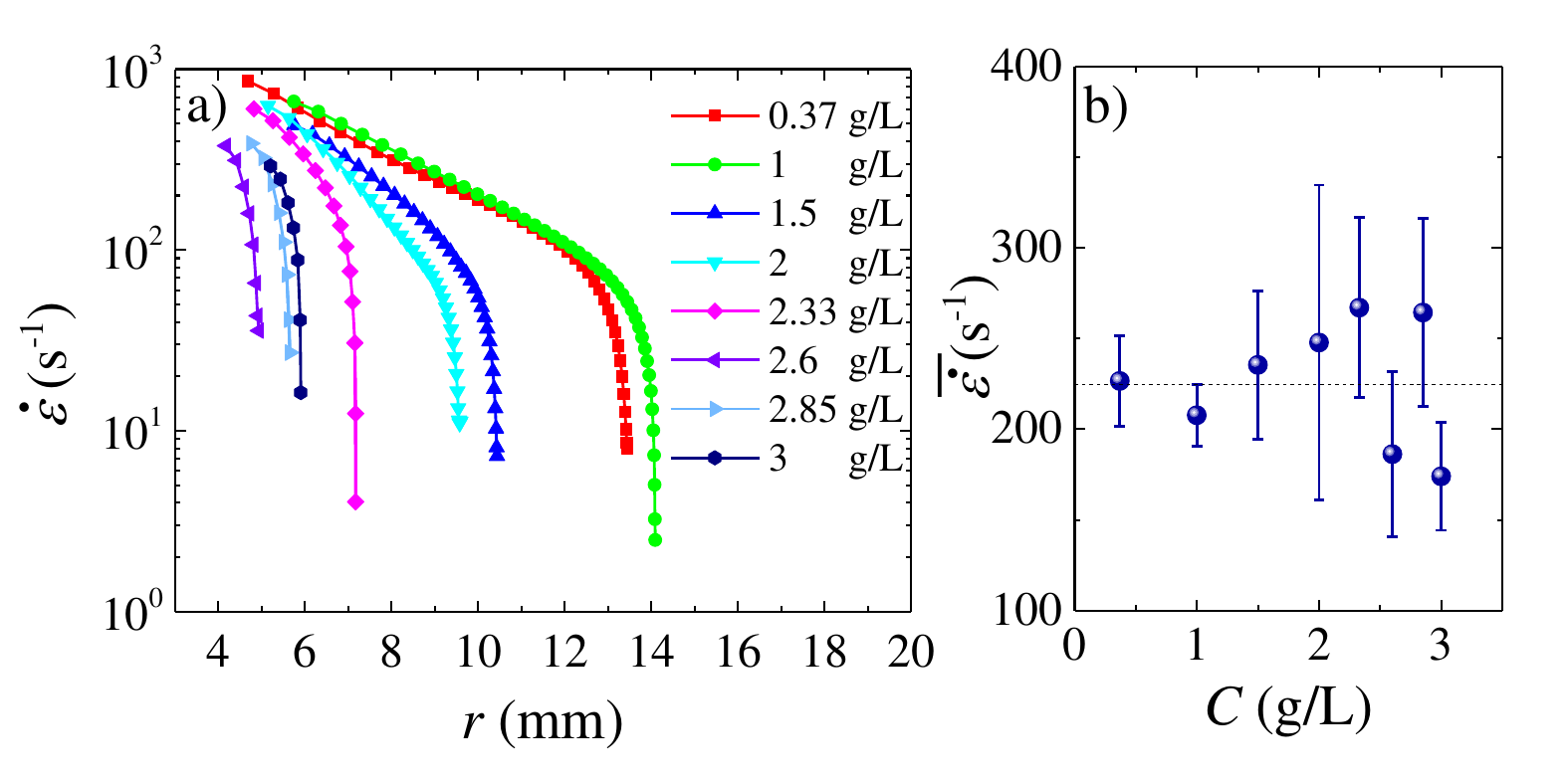}}
\caption{(a) Evolution of the biaxial extensional strain rate as a function of the radius of the sheet during its expansion for EHUT samples with different concentrations, as indicated in the legend. (b) Average strain rate calculated according to Eq.\ref{eqn:strainrate} as a function of concentration. Error bars represent the standard deviation of three different experiments.}
\label{figure:12}
\end{figure}

In the following, we will verify our assumption that the biaxial extensional dissipation is the relevant source of viscous dissipation in the present experimental conditions. First, we validate the rheological origin of $\eta\mathrm{^{eff}_B}$.
The effective biaxial extensional viscosity, $\eta\mathrm{^{eff}_B}$, is expected to depend on the strain rate. We use here the same approach as the one developed in \citep{Louhichi2020} to extract a relevant strain rate for the impact experiment. The time-dependent biaxial extensional  strain rate in the expansion regime (Figure \ref{figure:12}) is averaged as:

\begin{equation}
\label{eqn:strainrate}
\bar{\dot{\varepsilon}}=\frac{\int_0^{R_{\rm{max}}}r\dot{\varepsilon}dr}{\int_0^{R_{\rm{max}}}rdr}
\end{equation}

 Although the average strain rate (Figure \ref{figure:12}) does not vary significantly  with sample concentration ($\bar{\dot{\varepsilon}}_{\rm{Mean}}=225 \pm 35\,\mathrm{s^{-1}}$), in the following we use the computed average strain rate ($\bar{\dot{\varepsilon}}$) for each concentration.

\begin{figure}

\centerline{\includegraphics[width=8.3 cm]{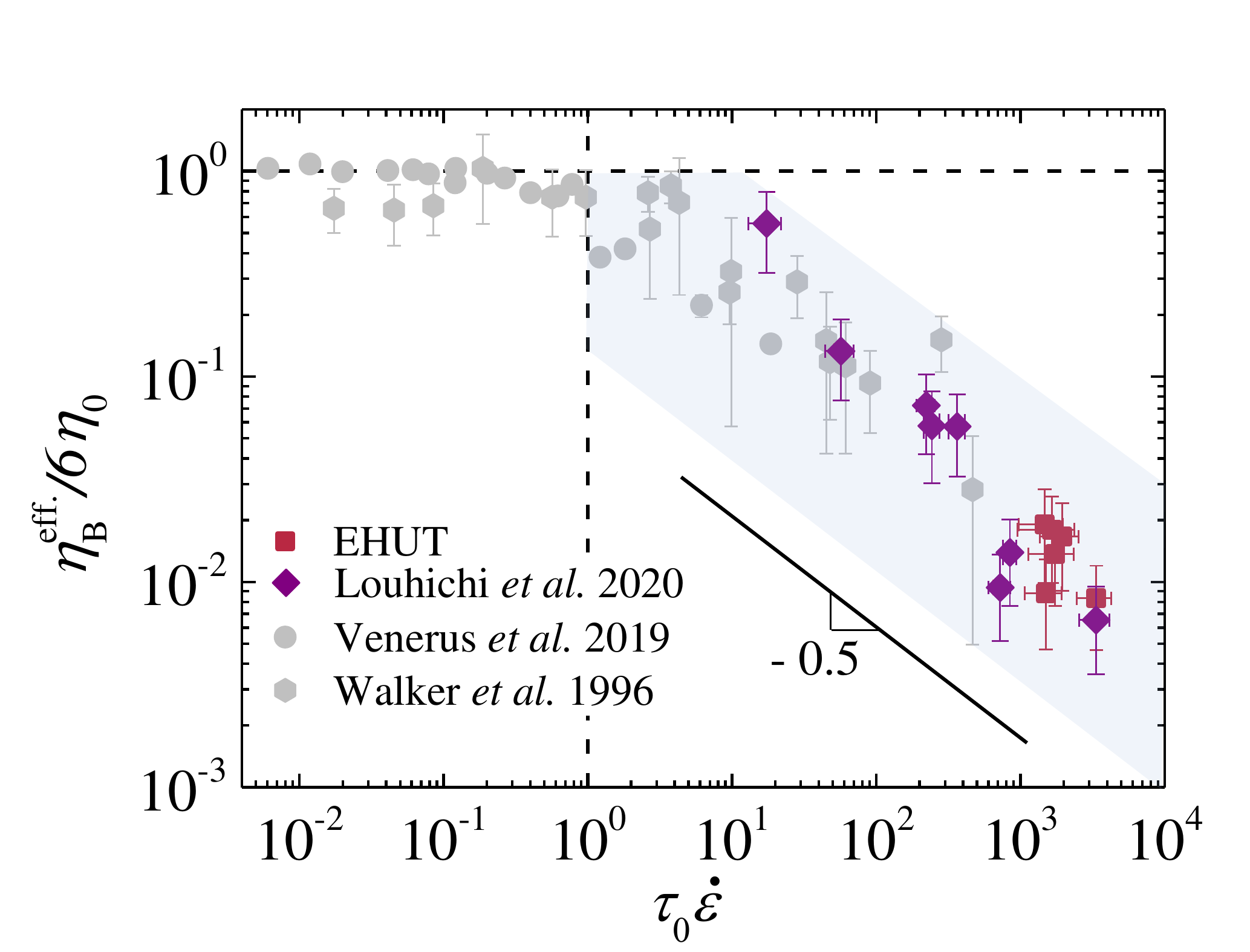}}

\caption{Effective thinning biaxial extensional viscosity, extracted from the manual shift of the data from Figure \ref{figure:11} in order to collapse with the data for Newtonian samples, normalized by the plateau biaxial extensional viscosity as a function of the Weissenberg number for the different EHUT solutions studied here (red squares), together with literature data as indicated in the legend. A unique master biaxial flow curve is obtained with a thinning exponent of $ -0.5 $ in agreement with theoretical predictions (see text). The grey zone is the region of data uncertainties}.

\label{figure:13}
\end{figure}

We report in Figure \ref{figure:13} the variation of the normalized effective biaxial extensional viscosity, $\eta\mathrm{^{eff}_B}/\eta^0_{\mathrm{B}}$, of EHUT supramolecular polymers as a function of the effective Weissenberg number, Wi$^{\mathrm{eff}}$. The latter is defined as: Wi$^{\mathrm{eff}}=\tau_0\bar{\dot{\varepsilon}}$, where $ \tau_0$ is the relaxation time obtained from linear shear rheology measurements (Figure \ref{figure:7}) and $\bar{\dot{\varepsilon}}$ is computed using Eq.\ref{eqn:strainrate} (Figure \ref{figure:12}).

On the same plot, we report previous experimental data for normalized biaxial extensional thinning viscosities measured for different viscoelastic fluids using different set-ups \citep{Walker1996,Venerus2019}. We also report PEO solutions data impacted under cold Leidenfrost conditions from our previous work \citep{Louhichi2020}. The EHUT experimental data follow remarkably well the experimental results  reported in the literature for stationary biaxial extensional deformation flows of different systems (surfactant wormlike micelles in \citep{Walker1996}, polystyrene solutions in \citep{Venerus2019} and PEO solutions in \citep{Louhichi2020}). At low Weissenberg numbers (Wi$= \dot{\varepsilon} \tau_0<1$), the biaxial extensional viscosity is rate independent and reaches the Newtonian limit: $\eta_{\mathrm{B}}=\eta_{\mathrm{B}}^{0}=6\eta_0$. By contrast, when Wi$>1$, the biaxial extensional viscosity decreases with the strain rate as $\eta_{\mathrm{B}}\sim6\eta_0(\dot{\epsilon}\tau_0)^{-0.5}$ with a thinning exponent of $-0.5$ as validated by theoretical predictions in the framework of the classical tube based model \citep{Marrucci2004}.

    \paragraph{Comparison between shear and biaxial extensional viscous dissipation}
    
After assessing $\eta_\mathrm{S}^\mathrm{eff}$ and validating the rheological origin of $\eta\mathrm{^{eff}_{B}}$, we can now compute and compare the two viscous dissipation processes for the supramolecular polymers. In order to account for the rate-dependent viscosity in the shear dissipation energy of non-Newtonian fluids, we have replaced, in Eqs. \ref{eqn:ESH1} and \ref{eqn:ESdelta1}, $\eta_S$ by $\eta_\mathrm{S}^\mathrm{eff}$. On the other hand, we replaced, in Eq. \ref{eqn:EB}, $\eta_{\mathrm{B}}$ by $\eta_{\mathrm{B}}^\textrm{eff}$, to evaluate the biaxial extensional viscous energy.

 Results are reported on Figure \ref{figure:6}. The dissipation energies have been normalized by the initial kinetic energy of the drop $E_\mathrm{K}=\frac{1}{2}mv_0^2$ and plotted as a function of the zero-shear rate Ohnesorge number. We observe that the biaxial extensional viscous dissipation energy is larger by two order of magnitude than the shear dissipation energy (for both limits in the calculation of the shear dissipation energy) despite the fact that the zero shear viscosity of the less concentrated complex fluid sample (about 10 Pa s) is two orders of magnitude larger than the more viscous Newtonian samples we have investigated (about 0.2 Pa s). Note that in the same experimental conditions a Newtonian viscous sample with a viscosity of 10 Pa s will not expand outside of the target after impact.

\section{Conclusions}

Drop impact experiments on a small solid target have been performed with Newtonian fluids and solutions of entangled supramolecular polymers as rheo-thinning viscoelastic fluids. Upon impact on the surface, a drop expands into a sheet. A part of the expanding sheet is always at intimate contact with the target while the rest of the sheet freely expands in the air. We have measured the maximum expansion of sheets for the two classes of fluids. We have quantitatively assessed the energy dissipation due to shear and biaxial extensional deformations, and have highlighted the distinctive roles of shear and biaxial extensional viscosities. We have shown that the expansion process is controlled by a combination of shear (on the target) and biaxial extensional (in the air) deformations. For Newtonian fluids, the two sources of dissipation remain of the same order of magnitude. In sharp contrast to the findings for Newtonian fluids, for the rheo-thinning viscoelastic fluids, the dominant source of viscous dissipation is the biaxial extensional deformation during sheet expansion, and consequently the biaxial thinning extensional viscosity is found to control the maximum expansion of the sheets. The physical reason is that the thinning behavior of the fluid strongly depends on the flow type, since nonlinear viscosities scale differently with the relevant deformation rate. In other words, polymeric fluids exhibit a different thinning behaviour in shear and in biaxial extension, justifying the distinction we made between the two viscous dissipation modes. On the contrary, for Newtonian fluids both viscosities are proportional to each other, thus the distinction of the nature of the viscous dissipation is less important. We have also shown that the ratio between the two dissipation modes strongly depends on the size of the target. It would therefore be interesting to study the contribution of the shear dissipation at different drop to target ratios. This will be the task of a future work.

\begin{acknowledgments}
This work was financially supported by the EU (Marie Sklodowska Curie) ITN Supolen GA $N^\circ.607937$, the labex NUMEV (ANR-10-LAB-20) and the H2020 Program (Marie Curie Actions) of the European Commission's Innovative Training Networks (ITN) (H2020-MSCA-ITN-2017) under DoDyNet REA Grant Agreement (GA) $N^\circ.765811$.

\end{acknowledgments}

\section*{data availability}
The data that support the findings of this study are available from the corresponding author upon reasonable request.

\bibliography{POF21-AR-02279_revision}

\end{document}